\input harvmac



\font\cmss=cmss10 \font\cmsss=cmss10 at 7pt
\def\rlx{\relax\leavevmode}

\def\RR{\relax{\rm I\kern-.18em R}}
\def\IZ{\relax\ifmmode\mathchoice
{\hbox{\cmss Z\kern-.4em Z}}{\hbox{\cmss Z\kern-.4em Z}}
{\lower.9pt\hbox{\cmsss Z\kern-.4em Z}} {\lower1.2pt\hbox{\cmsss
Z\kern-.4em Z}}\else{\cmss Z\kern-.4em Z}\fi}
\def\ZZ{\rlx\leavevmode\ifmmode\mathchoice{\hbox{\cmss Z\kern-.4em Z}}
 {\hbox{\cmss Z\kern-.4em Z}}{\lower.9pt\hbox{\cmsss Z\kern-.36em Z}}
 {\lower1.2pt\hbox{\cmsss Z\kern-.36em Z}}\else{\cmss Z\kern-.4em
 Z}\fi}
\def\frac#1#2{{#1 \over #2}}
\def\gs{{g'}}
\def\gsi{{g'^{-1}}}

\def\B{{\cal B}}
\def\Bs{{\cal B}^*}
\def\CL{{\cal L}}

\def\tR{{\tilde R}}

\def\JK{J_{K3}}
\def\a{\alpha}
\def\b{\beta}
\def\g{\gamma}
\def\d{\delta}
\def\e{\epsilon}

\def\th{\theta}

\def\t{\tau}

\def\p{\phi}

\def\G{\Gamma}

\def\g{\gamma}
\def\O{\Omega}
\def\La{\Lambda}
\def\om{\omega}

\def\tom{{\tilde \omega}}

\def\bd{{\bar d}}

\def\w{\wedge}

\def\tbe{{\tilde \beta}}
\def\pa{\partial}
\def\bpa{{\bar \partial}}
\def\dg{\dagger}
\def\o{\over}

\def\bbe{{\bar \beta}}

\def\bth{{\bar \theta}}

\def\bj{{\bar j}}

\font\tenmsx=msam10
\font\sevenmsx=msam7
\font\fivemsx=msam5
\newfam\msxfam
\textfont\msxfam=\tenmsx  \scriptfont\msxfam=\sevenmsx
  \scriptscriptfont\msxfam=\fivemsx
\def\hexnumber@#1{\ifcase#1 0\or1\or2\or3\or4\or5\or6\or7\or8\or9\or%
A\or B\or C\or D\or E\or F\fi }
\edef\msx@{\hexnumber@\msxfam}
\def\llcorner{\,\delimiter"4\msx@78\msx@78 \,}

\def\ap{\alpha'}

\def\rd{{ d}}

\def\ba{{\bar a}}
\def\bb{{\bar b}}
\def\bc{{\bar c}}
\def\bd{{\bar d}}

\def\bg{{\bar g}}
\def\bh{{\bar h}}

\def\bz{{\bar z}}


\lref\Strom{ A.~Strominger, ``Superstrings with torsion,'' Nucl.\
Phys.\ B {\bf 274}, 253 (1986). }

\lref\dewit{B.~de Wit, D.~J.~Smit and N.~D.~Hari Dass, ``Residual
supersymmetry of compactified d=10 supergravity,'' Nucl.\ Phys.\ B
{\bf 283}, 165 (1987).}

\lref\caho{P.~Candelas, G.~T.~Horowitz, A.~Strominger and
E.~Witten,``Vacuum Configurations For Superstrings,''
  Nucl.\ Phys.\ B {\bf 258}, 46 (1985).}

\lref\Strominger{A.~Strominger, ``Special geometry,''
  Commun.\ Math.\ Phys.\  {\bf 133}, 163 (1990).}

\lref\FY{J.-X.~Fu and S.-T.~Yau, ``The theory of superstring with
flux on non-K\"ahler manifolds and the complex Monge-Amp\'ere
equation,'' [arXiv:hep-th/0604063].
}

\lref\BBFTY{
  K.~Becker, M.~Becker, J.-X.~Fu, L.-S.~Tseng and S.-T.~Yau,
   ``Anomaly cancellation and smooth non-K\"ahler solutions in heterotic string theory,''
  Nucl.\ Phys.\ B {\bf 751}, 108 (2006)
  [arXiv:hep-th/0604137].
} \lref\adel{
  A.~Adams, M.~Ernebjerg and J.~M.~Lapan,
  ``Linear models for flux vacua,''
  [arXiv:hep-th/0611084].
} \lref\kimyi{
  S.~Kim and P.~Yi,
  ``A heterotic flux background and calibrated five-branes,'' JHEP {\bf 0611}, 040
  (2006) [arXiv:hep-th/0607091].
} \lref\kiyi{
  T.~Kimura and P.~Yi,
  ``Comments on heterotic flux compactifications,''
  JHEP {\bf 0607}, 030 (2006)
  [arXiv:hep-th/0605247].
} \lref\cylap{
  M.~Cyrier and J.~M.~Lapan,
``Towards the massless spectrum of non-Kaehler heterotic
compactifications,''
  [arXiv:hep-th/0605131].
} \lref\LiYau{J.~Li and S.~T.~Yau, ``The existence of supersymmetric
string theory with torsion,'' J.\ Differential\ Geom.\  {\bf 70},
143 (2005) [arXiv:hep-th/0411136].
}


\lref\BC{R.~Bott and S.~S.~Chern, "Hermitian vector bundles and
the equidisstribution of the zeroses of their holomorphic
sections," Acta.\ Math.\ {\bf 114} 71 (1965).}

\lref\liyau{J.~Li and S.-T.~Yau, ``Hermitian-Yang-Mills connection
on non-K\"ahler manifolds,'' in {\it Mathematical aspects of string
theory}, World Scientific Publ., S.-T.~Yau, editor; London 560
(1987).}

\lref\Lubke{M.~L\"ubke and A.~Teleman, {\it The Kobayashi-Hitchin
correspondence}, World Scientific, River Edge, NJ (1995).}

\lref\Mukai{S.~Mukai, ``Moduli of vector bundles on K3 surfaces, and
symplectic manifolds,'' Sugaku Expositions, {\bf 1}, 139 (1988).}

\lref\berroo{ E.~A.~Bergshoeff and M.~de Roo, { ``The quartic
effective action of the heterotic string and supersymmetry,''}
Nucl.\ Phys.\ B {\bf 328}, 439 (1989).}

\lref\strom{A.~Strominger, {``Superstrings with torsion,''}
Nucl.\ Phys.\ B {\bf 274}, 253 (1986).}

\lref\Hullnk{C.~M.~Hull,
 ``Superstring compactifications with torsion and space-time supersymmetry,''
 in {\it  1st Torino meeting on superunification and extra dimensions, September 1985, Torino, Italy}, World Scientific,  R.~D'Auria and D.~Fre, editors; Singapore, 347 (1986).}

\lref\dewit {B.~de Wit, D.~J.~Smit and N.~D.~Hari Dass, {
``Residual supersymmetry of compactified d=10 supergravity''} Nucl.\
Phys.\ B {\bf 283}, 165 (1987).}

\lref\dbeck{K.~Dasgupta, G.~Rajesh and S.~Sethi, ``M theory,
orientifolds and G-flux,''  JHEP {\bf 9908}, 023 (1999)
  [arXiv:hep-th/9908088];
K.~Becker and K.~Dasgupta, ``Heterotic strings with torsion,''
  JHEP {\bf 0211}, 006 (2002)
  [arXiv:hep-th/0209077].
}

\lref\beck{K.~Becker, M.~Becker, K.~Dasgupta and P.~S.~Green, {
``Compactifications of heterotic theory on non-K\"ahler complex
manifolds I,''}JHEP {\bf 0304}, 007 (2003),
[arXiv:hep-th/0301161]; 
K.~Becker, M.~Becker, P.~S.~Green, K.~Dasgupta and E.~Sharpe,{
``Compactifications of heterotic strings on non-K\"ahler complex
manifolds II,''} Nucl.\ Phys.\ B {\bf 678}, 19 (2004),
[arXiv:hep-th/0310058].}

\lref\loca{G.~Lopes Cardoso, G.~Curio, G.~Dall'Agata, D.~Lust,
P.~Manousselis and G.~Zoupanos, { ``Non-K\"ahler string
backgrounds and their five torsion classes,''} Nucl.\ Phys.\ B
{\bf 652}, 5 (2003), [arXiv:hep-th/0211118].}

\lref\douglas{M.~R.~Douglas and S.~Kachru, { ``Flux
compactification,''} [arXiv:hep-th/0610102].}

\lref\grana{M.~Grana, ``Flux compactifications in string theory: A
comprehensive review,'' Phys.\ Rept.\  {\bf 423}, 91 (2006)
[arXiv:hep-th/0509003].}

\lref\lolu{G.~Lopes Cardoso, G.~Curio, G.~Dall'Agata and D.~Lust,
{\it ``BPS action and superpotential for heterotic string
compactifications with fluxes,''} JHEP {\bf 0310}, 004 (2003)
[arXiv:hep-th/0306088].}

\lref\caos{P.~Candelas and X.~de la Ossa, { ``Moduli space of
Calabi-Yau manifolds,''}, Nucl.\ Phys.\ B {\bf 355}, 455 (1991).}

\lref\lika{K.~Becker and L.~S.~Tseng, { ``Heterotic flux
compactifications and their moduli,''}Nucl.\ Phys.\ B {\bf 741}, 162
(2006), [arXiv:hep-th/0509131].}

\lref\Candelas{P.~Candelas, ``Yukawa couplings between (2,1)
forms,''
  Nucl.\ Phys.\ B {\bf 298}, 458 (1988);
 P.~Candelas and X.~de la Ossa,  "Moduli space of Calabi-Yau manifolds,"
   Nucl.\ Phys.\ B {\bf 355}, 455 (1991).
}

\lref\gmpw{
J.~P.~Gauntlett, D.~Martelli, S.~Pakis and D.~Waldram,
  Commun.\ Math.\ Phys.\  {\bf 247}, 421 (2004)
  [arXiv:hep-th/0205050].}

\lref\locu{
  G.~Lopes Cardoso, G.~Curio, G.~Dall'Agata and D.~Lust, ``BPS action and superpotential
  for heterotic string compactifications  with fluxes,'' JHEP {\bf 0310}, 004
  (2003), [arXiv:hep-th/0306088].}

\lref\ovrut{V.~Braun, Y.~H.~He, B.~A.~Ovrut and T.~Pantev,
  ``A heterotic standard model,''
  Phys.\ Lett.\ B {\bf 618}, 252 (2005)
  [arXiv:hep-th/0501070].}

\lref\bouchard{V.~Bouchard and R.~Donagi, ``An SU(5) heterotic
standard model,'' Phys.\ Lett.\ B {\bf 633}, 783 (2006),
[arXiv:hep-th/0512149].}

\lref\herman{H.~Verlinde and M.~Wijnholt, ``Building the standard
model on a D3-brane,'' [arXiv:hep-th/0508089].}

\lref\Mukai{S.~Mukai, ``Moduli of vector bundles on K3 surfaces,
and symplectic manifolds,'' Sugaku Expositions, {\bf 1}, 139
(1988).}

\lref\cvetic{ R.~Blumenhagen, M.~Cvetic, P.~Langacker and G.~Shiu,
``Toward realistic intersecting D-brane models,'' Ann.\ Rev.\ Nucl.\
Part.\ Sci.\  {\bf 55}, 71 (2005)
  [arXiv:hep-th/0502005].}

\lref\witten{D.~J.~Gross and E.~Witten, ``Superstring
modifications of Einstein's equations,'' Nucl.\ Phys.\ B {\bf
277}, 1 (1986).}

\lref\sen{D.~Nemeschansky and A.~Sen, ``Conformal invariance of
supersymmetric simga models on Calabi-Yau manifolds,'' Phys.\
Lett.\ B {\bf 178}, 365 (1986).}

\lref\Hull{C.~M.~Hull, ``Actions for (2,1) sigma models and
strings,'' Nucl.\ Phys.\ B {\bf 509}, 252 (1998)
[arXiv:hep-th/9702067].}

\lref\BBDP{
  K.~Becker, M.~Becker, K.~Dasgupta and S.~Prokushkin,
  ``Properties of heterotic vacua from superpotentials,''
  Nucl.\ Phys.\ B {\bf 666}, 144 (2003)
  [arXiv:hep-th/0304001].
}

\lref\BT{
  K.~Becker and L.-S.~Tseng,
  ``Heterotic flux compactifications and their moduli,''
  Nucl.\ Phys.\ B {\bf 741}, 162 (2006)
  [arXiv:hep-th/0509131].
}


\Title{ {\vbox{ \rightline{\hbox{hep-th/0612290}}
\rightline{\hbox{HUTP-06/A0046}} }}} {\vbox{ \hbox{\centerline{
Moduli Space of Torsional Manifolds}}\hbox{} \hbox{\centerline{}}
}}

\centerline{Melanie Becker$^{1}$, ~Li-Sheng
Tseng$^{2,3}$~and~Shing-Tung Yau$^3$}
\bigskip
\bigskip
\centerline{$^1$ \it George P. and Cynthia W. Mitchell Institute
for Fundamental Physics} \centerline{\it Texas A \& M University,
College Station, TX 77843, USA}
\smallskip
\centerline{$^2$\it Jefferson Physical Laboratory, Harvard University,
Cambridge, MA 02138, USA}
\smallskip
\centerline{$^3$\it Department of Mathematics, Harvard University,
Cambridge, MA 02138, USA}

\bigskip

\bigskip

\bigskip

\centerline{\bf Abstract}

\bigskip
We characterize the geometric moduli of non-K\"ahler manifolds
with torsion.  Heterotic supersymmetric flux compactifications
require that the six-dimensional internal manifold be balanced,
the gauge bundle be hermitian Yang-Mills, and also the anomaly
cancellation be satisfied. We perform the linearized variation of
these constraints to derive the defining equations for the local
moduli. We explicitly determine the metric deformations of the
smooth flux solution corresponding to a torus bundle over $K3$.

\bigskip
\baselineskip 18pt
\bigskip
\noindent

\Date{December, 2006}
\newsec{Introduction}
Ever since the discovery of Calabi--Yau compactifications \caho,
string theorists have tried to make the connection to the minimal
supersymmetric standard model (MSSM) and grand unified theories
(GUT). This turned out to be a difficult problem, as many times
``exotic particles'' appear along the way. These are particles
that play no role in the current version of the MSSM\foot{It is,
of course, possible that additional particles not known at present
might be discovered, leading to an extension of the MSSM.}.
Recently \refs{\ovrut,\bouchard} have made a rather interesting
proposal for three generation models without exotics in the
context of Calabi--Yau compactifications of the heterotic
string.\foot{String duality implies that in principle one could
get realistic models in the context of type II theories. A
concrete proposal has been made recently in terms of a D3-brane in
the presence of a $dP_8$ singularity \refs{\herman}.
Alternatively, one could use intersecting D-brane models. For a
review see \refs{\cvetic}.}

Even though these models have some rather interesting features, it
is not possible to predict with them the values of the coupling
constants of the standard model, because compactifications on
conventional Calabi--Yau compactifications lead to unfixed moduli,
and therefore additional massless scalars. This issue can
only be addressed in the context of flux compactifications, which
are known to lift the moduli \refs{\douglas, \grana}.

If flux compactifications are considered in the context of the
heterotic theory, the resulting internal geometry is a
non-K\"ahler manifold with torsion \refs{\Hullnk, \strom,\dewit}. Simple
examples of such compactifications were constructed in
\refs{\dbeck,\beck} in the orbifold limit and a smooth
compactification was constructed in \refs{\FY, \BBFTY} in terms of
a $T^2$ bundle over $K3$. See \refs{\cylap, \kiyi, \kimyi, \adel}
for some related works. It would be extremely exciting to
construct a torsional manifold with all the features of the MSSM.
At present, we are not yet at such a state. Many properties of
Calabi--Yau manifolds are not shared by non-K\"ahler manifolds
with torsion, so that well known aspects of Calabi--Yau manifolds
need to be rederived for these manifolds.

One of the important open questions is to understand how to
characterize the scalar massless fields, in other words, the
moduli space of heterotic flux compactifications. We investigate
this question by analyzing the local moduli space emerging in such
compactifications from a spacetime approach. A massless scalar
field in the effective four-dimensional theory emerges for each
independent modulus of the background geometry. Thus, the
dimension of the moduli space corresponds to the number of
massless scalar fields in the theory. In our analysis, we restrict
to supersymmetric deformations, as we expect the analysis of the
supersymmetry constraints to be easier than the analysis of the
equations of motion. While the later equations are corrected by
$R^2$ terms, the form of the supersymmetry transformations is not
modified to $R^2$ order, as long as the heterotic anomaly
cancellation condition is imposed \berroo.  That a solution of
both the supersymmetry constraints and the modified Bianchi
identity is also a solution to the equations of motion has been
shown in \refs{\gmpw, \locu}.

Unlike the Calabi--Yau case, the supersymmetry constraint
equations in general non-linearly couple the various fields and
thus the analysis even at the linearized variation level is
non-trivial. As an example of our general analysis, we shall give
the description of the scalar metric moduli for the smooth
solution of a $T^2$ bundle over $K3$ presented in \refs{\FY,
\BBFTY}. It is an interesting question to understand whether the
massless moduli found in our approach are lifted by higher order
terms in the low energy effective action. For conventional
Calabi--Yau compactifications it is known that moduli fields
appearing in the leading order equations will remain massless even
if higher order corrections are taken into account \refs{\witten,
\sen}. In our case, such an analysis has not been performed yet
from the spacetime point of view, though the question can be
answered from the world-sheet approach recently developed in
\refs{\adel}. In this work, a gauged linear sigma model was
constructed which in the IR flows to an interacting conformal
field theory. The analysis of the linear model indicates that
massless fields emerging at leading order in $\alpha'$ will remain
massless, even if corrections to the spacetime action are taken
into account.

This paper is organized as follows. In section 2, we perform the
linear variation of the supersymmetry constraints.  In section 3,
we analyze the variation of the $T^2$ bundle over $K3$ solution
and discuss its local moduli space. In section 4, conclusions and
future directions are presented. In the appendix, we clarify some
of mathematical notations that we used.


\newsec{Determining equations for the moduli fields}

The non-K\"ahler manifolds with torsion ${\cal M}$ that we are
interested in are complex manifolds described in terms of a
hermitian form which is related to the metric
\eqn\hform{J=i\,g_{a\bar b}\,dz^a\wedge d\bz^{\bar b}~,} 
and a no-where vanishing holomorphic three-form
\eqn\oform{d\,\O = 0~,}
satisfying $J\w \O =0$.  The geometry can be deformed by either 
deforming the hermitian form or
deforming the complex structure of ${\cal M}$. We are interested
in deformations that preserve the supersymmetry constraints as
well as the anomaly cancellation condition.

${\cal N}=1$ supersymmetry for heterotic flux compactifications to
four spacetime dimensions imposes three conditions: the internal
geometry has to be conformally balanced, the gauge bundle
satisfies the hermitian Yang-Mills equation, and the $H$-flux
satisfies the anomaly cancellation condition. Explicitly, they are
\refs{\FY, \BBFTY}
\eqn\balance{\rd(\|\Omega\|_J\, J\w J)=0~,}
\eqn\hermitym{F^{(2,0)}=F^{(0,2)}=0~,\quad F_{mn}J^{mn}=0~,}
\eqn\anomcan{2i\, \pa\bpa J = {\ap \o 4} [{\rm tr} (R\w R) - {\rm
tr}(F\w F)]~.}

Above, we have replaced the two standard background
fields - the three-form $H$ and the dilaton field $\p$ - with the
required supersymmetric relations
\eqn\Hdef{H=i(\bpa-\pa) J~,}
\eqn\Omdef{\|\Omega\|_J=e^{-2(\p+\p_0)}~.}
Doing so allows us to consider the constraint equations solely
in terms of the geometrical data $(J, \O)$ and the gauge bundle.

Deformations of the metric that are of pure type, i.e. $(0,2)$ or
$(2,0)$, describe deformations of the complex structure
\eqn\defcs{\Omega_{ab}{}^\bd\, \delta g_{\bar d \bar c}\,\, dz^a
\wedge dz^b \wedge d\bz^{\bar c}~,} while deformations of mixed
type, i.e. of type $(1,1)$, describe deformations of the hermitian
form \eqn\defhf{i\, \d g_{a\bar b}\,\,dz^a\wedge d\bz^{\bar b}~.}

We analyze below the linear variation of the three constraint
equations \balance-\anomcan\ with respect to a background
solution. For simplicity, we shall keep the complex structure of
the six-dimensional internal geometry fixed. For the moduli space
of Calabi-Yau compactifications, it turns out that the K\"ahler
and complex structure deformations decouple from one another
\Candelas. It would be interesting to determine whether some
decoupling still persists in the non-K\"ahler case and more
generally how the hermitian and complex structure deformations are
coupled. We will leave this more general analysis for future work.


\subsec{Conformally balanced condition}

We consider the linear variation of the conformally balanced
condition \balance. We shall vary the metric or hermitian form
$J_{a\bb}=ig_{a\bb}$ while holding fixed the complex structure.  Let
\eqn\metvar{J_{a\bar b}' = J_{a\bar b} +\d J_{a\bar b}~,} then we
have to first order in $\d J$ \eqn\Jsvar{J'\w J' =  J\w J + 2 J \w
\d J~,} \eqn\Omvar{\eqalign{\|\Omega\|^2_{J'} &= \frac{|g_{a\bb}|}
{|g'_{a\bb}|}\|\Omega\|^2_{J}=\frac{|g_{a\bb}|}{|g_{a\bb}| (1+
g^{c\bd} \d g_{c\bd})}\|\Omega\|^2_{J},\cr &=
 (1 - g^{c\bd} \d g_{c\bd})\|\Omega\|^2_{J} ~.} }
Note that \Omdef\ with \Omvar\ imply the dilaton variation
\eqn\dilvar{\d \p = \frac{1}{4}\, g^{a\bb} \d g_{a\bb}  =
\frac{1}{8}\,J^{mn} \d J_{mn} ~.} The linear variation of the
conformally balanced condition can be written as
\eqn\confbalv{d\left(\|\Omega\|_{J'}\, J'\w
J'\right)=d\left(\|\Omega\|_J\, J\w J+ 2\,\d\rho\right)=0~,} where
$\d \rho$ is a four-form given by \eqn\confbalvv{ \d\rho =
\|\O\|_J\left[J\w \d J - \frac{1}{8} (J \w J) J^{mn}\d
J_{mn}\right]~.}

We can invert \confbalvv\ and express $\d J$ in terms of $\d\rho$.
To do this, we note that any $(2,2)$-form, $\om_4$, can be
Lefschetz decomposed as follows \eqn\Ldecomp{\om_4= L \La\, \om_4
- \frac{1}{4} L^2 \La^2 \om_4,} where the Lefschetz operator $L$
and its adjoint $\La$ have the following action on exterior forms
\eqn\LLact{\eqalign{ L:&~~ \om \to J \w \om, \cr \La:&~~ \om \to J
\llcorner  \om.}} Comparing \confbalvv\ with \Ldecomp, we find the
relation \eqn\Jrho{ \d J_{mn} = \frac{1}{2\|\O\|_J}\, \d
\rho_{mnrs} \,J^{rs}.}

From the linear variation of equation \confbalv, we observe that
the allowed deformations (i.e. which preserve the conformally
balanced condition) satisfy $d\, \d \rho =0$. Equation \Jrho\
implies that any variation of the hermitian metric can be
expressed in terms of a variation by a closed $(2,2)$-form.
Equivalently, we can also express the linear variation condition
directly for the hermitian metric as \eqn\confbalt{d^*[\d J' -
\frac{1}{4}J (J^{mn}\d J'_{mn})] = 0~, } where $\d J' =
\|\Omega\|_J \d J$.

Note that $\d J$ variations that are equivalent to a coordinate
transformation (i.e. a diffeomorphism) are physically unobservable
and must therefore be quotient out. Under an infinitesimal
coordinate transformation \eqn\coortrans{y'^m = y^m + v^m(y)~,}
the variation of a $p$-form $\om_p$ is given by the Lie derivative
\eqn\omvar{\d \om_p = - \CL_v\,\om_p = - [i_v(d\,\om_p) + d(i_v\,
\om_p)]~, } where $v=v^m\pa_m$ is a vector field and $i_v$ denotes
the interior product.  For the conformally balanced four-form, a
coordinate transformation results in \eqn\balcoor{\CL_v
(\|\Omega\|_J\, J\w J)  = d \left[i_v (\|\Omega\|_J\, J\w
J)\right]~. } We can thus identify, as physically not relevant,
$\d \rho$ variations that are exterior derivatives of a
non-primitive three-form \eqn\rhodiffeo{ \d \rho \sim d (\|\O\|_J
\b \w J),} where $\b_m = v^n J_{nm}$. Using \Jrho, this
corresponds to deformations of the hermitian form \eqn\Jdiffeo{\d
J \sim \frac{1}{\|\O\|_J} \La\, d (\|\O\|_J \b \w J)~.}

Let us now interpret the content of the above variation formulas.
By the identification of \Jrho, variations of the hermitian metric
that preserve the conformally balanced condition can be
parametrized by closed $(2,2)$-forms.  Moreover, modding out by
diffeomorphisms results in the cohomology\foot{Note that complex
structures are also defined up to diffeomorphism. So any
diffeomorphism generated by a real vector field will keep the
complex structure in the same equivalence class.}
\eqn\balcohfour{\frac{{\rm ker}(d) \cap \Lambda^{2,2} } {d(\b \w
J)}~.} Thus, the space of conformally balanced metrics is
equivalent to the space of closed $(2,2)$-forms modded out by
those which are exterior derivatives of a non-primitive
three-form.  But notice that exact forms which are exterior
derivative of a primitive three-form are not quotient out. Hence,
if there exists such a primitive three-form, $\om_3^0$, then the
space of balanced metrics is infinite dimensional.  This is
because $d(f\om_3^0)$ where $f$ is any real function would be
closed but not modded out.

The cohomology of \balcohfour\ can also be expressed directly in
terms of $(1,1)$-forms. From \confbalt, every co-closed
$(1,1)$-form defines a metric deformation preserving the
conformally balanced condition. To see this explicitly, we note
that any $(1,1)$-form can be Lefschetz decomposed as follows
\eqn\twodecomp{\eqalign{C_{mn}&=(C_0)_{mn}+\frac{1}{6}\, J_{mn}\,
J^{rs}C_{rs}\cr &\equiv (C_0)_{mn} +\frac{1}{3}\,J_{mn}\,C_\La~,}}
where $C_0$ denotes the primitive part and $C_\La=\frac{1}{2}
J^{rs}C_{rs}$ encodes the non-primitivity of $C_{mn}\,$.  We can
therefore re-express \confbalt\ as \eqn\confbalts{\eqalign{0&=
d^*(\d J' - \frac{1}{2} J \,\d J'_\La)\cr &= d^*(\d J'_0 -
\frac{1}{6} J\,\d J'_\La)\cr & = d^* C~,}} where we have defined a
new $(1,1)$-form $C=C_0 + \frac{1}{3} J\, C_\La$ with $C_0=\d
J'_0$ and $C_\La=-\frac{1}{2} \d J'_\La$.  Furthermore, variations
associated with diffeomorphisms can be written as
\eqn\diffeomorp{\d J' \sim \La d(\|\O\|_J \b\w J)~, } so that we
have \eqn\diffeomtwo{(\d J' - \frac{1}{2} J\d J'_\La) \sim
d^*(\tbe'\w J),} where $\tbe'_m = J_m{}^n \b_n \|\O\|_J$.
Equations \confbalts\ and \diffeomtwo\ together imply the
cohomology \eqn\balcohtwo{\frac{{\rm ker}(d^*) \cap \Lambda^{1,1}
}{d^*(\b \w J)}~.} Therefore, the local moduli space can also be
described as spanning all co-closed $(1,1)$-forms modulo those
which are $d^*$ of non-primitive three-forms. This space is
isomorphic to that of \balcohfour\ and is in general
infinite-dimensional. We have however yet to consider the two
other supersymmetry constraints. Imposing them, especially the
anomaly cancellation condition, will greatly reduce the number of
allowed deformations and render the moduli space
finite-dimensional.  This can be seen clearly in the $T^2$ bundle over $K3$ example
discussed in the next section.

Finally, let us point out that if we had taken into consideration
variations of the complex structure, then a $\d J$ variation will
in general include also a $(2,0)$ and a $(0,2)$ part.
Nevertheless, $J+ \d J$ must still be a $(1,1)$-form with respect
to the deformed complex structure as is required by supersymmetry.

\subsec{Hermitian Yang-Mills condition}

Any variation of the hermitian gauge connection with the complex
structure held fixed will preserve the holomorphic condition
$F^{(2,0)}=F^{(0,2)}=0$. As for the primitivity condition
$F_{mn}J^{mn}=0$, we shall vary its equivalent form \eqn\FJsvar{0=
\d (F \w J\w J) = \d F \w J^2 + 2 \,F \w J \w \d J~.} The hermitian
field strength $F$ can be written as \eqn\Fdef{F_{\ba b} = \bpa_\ba
A_b = \bpa_\ba( h^{\a\bbe}\pa_b h_{\bbe \g} ) =
 \bpa_\ba(\bh^{-1}\pa_b\bh ),}
where $\a, \bbe, \g$ are gauge indices and $\bh=h_{\bbe \a}$ is the
transpose of the hermitian metric on the gauge bundle. Under the
variation, $\bh'=\bh+\d \bh$, the gauge field varies as
\eqn\Avar{\eqalign{\d A=A'-A & = \bh^{-1} \pa(\d\bh) +
\d\bh^{-1}\pa\bh \cr & =
 \bh^{-1}\pa[\bh(\bh^{-1}\d\bh)] - \bh^{-1}\d\bh (\bh^{-1}\pa\bh) \cr &= \pa(\bh^{-1}\d\bh) +
  A (\bh^{-1}\d\bh) - (\bh^{-1}\d\bh) A\cr &\equiv D^A(\bh^{-1}\d\bh).}}
This implies that the field strength varies as $\d F =  \bpa
(D^A(\bh^{-1}\d\bh))$. Inserting into \FJsvar, we obtain
\eqn\FJsvars{0= \bpa (D^A(\bh^{-1}\d\bh) ) \w J^2 + 2\, F\w J\w \d
J~.} This gives the constraint relation between the variations of
the hermitian form and the gauge field. The pair $(\d J, \d h)$
will be further constrained when inserted into the anomaly
cancellation condition as we now show.

\subsec{Anomaly cancellation condition}

We can write the variation of the anomaly cancellation equation
as, \eqn\anomvar{2i \,\pa\bpa \,\d J = \frac{\ap}{2} \left( {\rm
tr} [R(g)\w \d R(g)] - {\rm tr}[F(h)\w \d F(h)]\right)~.} The left
hand side is a $\pa\bpa$ of a (1,1)-form, so we should write the
variation of the right hand side of the equation similarly.  With
the curvature defined using the hermitian connection, we can write
the variation using the Bott-Chern form \refs{\BC,\Hull}. For two
hermitian metrics $(g_1, g_0)$ that are smoothly connected by a
path parameterized by a parameter $t\in [0,1]$, the difference of
the first Pontraygin classes is given by the Bott-Chern form
\eqn\boch{{\rm tr}[R_1 \w R_1] - \,{\rm tr}[R_0 \w R_0] =
2i\,\pa\bpa BC_2(g_1, g_0)~,} where \eqn\bcform{BC_2(g_1,g_0) = 2i
\int_0^1 {\rm tr}[R_t\, \bg_t^{-1}{\dot \bg}_t] \, dt~~,}
and $\bg=g_{\ba b}$ denotes the transpose of the hermitian metric, the
dot denotes the derivative with respect to $t$, and the "tr" in
\bcform\ traces over only the holomorphic indices.\foot{Note that
the Bott-Chern form is defined only up to $\pa$ and $\bpa$ exact
terms.} We now use the Bott-Chern formula to obtain the variation.
Let \eqn\bcgg{g_t=g+ t\,(g'-g)= g + t\,\d g~,} where $t\in
[0,1]\,$ and in particular $g_0=g$ and $g_1=g'$. Then to first
order in $\d g$, we have
\eqn\rrvar{\d({\rm tr} [R\w R])= 2\,{\rm
tr} [R \w \d R] = -4\,\pa\bpa\,( {\rm tr} [R \,\bg^{-1}\d \bg]
)~,}
where the trace can be more simply written in components as
\eqn\tracrel{{\rm tr}[ R\,\bg^{-1}\d \bg]_{a\bb}= -i\,
R_{a\bb}{}^{c\bd}\d J_{c\bd}~.} With \rrvar, the linear variation
of the anomaly equation \anomvar\ becomes
\eqn\anomvarr{2i\,\pa\bpa \,\d J = -\ap \pa\bpa \left({\rm tr} [R \,\bg^{-1}\d \bg]
- {\rm tr} [F \,\bh^{-1}\d \bh] \right)~.}
By factoring out the $2i\pa\bpa$ derivatives, the anomaly condition
can be equivalently expressed as
\eqn\anomv{ \d J - i \frac{\a'}{2}\left( {\rm tr} [R
\,\bg^{-1}\d \bg] - {\rm tr} [F \,\bh^{-1}\d \bh]\right) = \g ~,}
where $\g$ is a $\pa\bpa$ closed $(1,1)$-form.

Note that for the special case where either the gauge bundle is
trivial (i.e. $F=0$) or $\d h=0$, there is a simple relationship
between $\d J$ and $\g$.  The anomaly variation with \tracrel\
inserted into \anomv\ becomes
\eqn\anomng{ \d J_{a\bb} - \frac{\a'}{2} R_{a\bb}{}^{c\bd}\d J_{c\bd} = \g_{a\bb}~.}
Grouping the two hermitian indices $(a\,\bb)$ as a single index, we can
solve for $\d J$ by inverting the above equation and obtain
\eqn\anoinvt{ \d J = (1 - M)^{-1} \g ~,} where the curvature is
encoded in the matrix $M_{a\bb}{}^{c\bd}=\frac{\a'}{2}
R_{a\bb}{}^{c\bd}$. As long as $(1 - M)$ is invertible, we see
that $\d J$ is parametrized by the space of $\pa\bpa$-closed
$(1,1)$-forms $\g$. Modding out by diffeomorphism equivalence, we
can obtain a cohomology associated with the anomaly equation of
the form
\eqn\anomcoh{\frac{{\rm ker}(\pa\bpa) \cap \Lambda^{1,1}}{\gamma_{di\!f\!f}}~,}
where
\eqn\gdiff{\gamma_{di\!f\!f} = (1-M)\, \d J_{di\!f\!f}=\frac{1}{\|\O\|_J}(1-M)\, \La\, d (\|\O\|_J \b \w J)~,}
and $\d J_{di\!f\!f}$ is the variation of the hermitian form
corresponding to diffeomorphism given in \Jdiffeo.

To summarize, we list the three linear variation conditions with
complex structure fixed. \eqn\lconda{d\big(\|\O\|_J[2J\w \d J -
\frac{1}{4} (J \w J) J^{mn}\d J_{mn}]\big) =0~,} \eqn\lcondb{\bpa
(D^A(\bh^{-1}\d\bh) ) \w J^2 + 2\, F\w J\w \d J =0 ~,}
\eqn\lcondc{\pa\bpa\Big(\d J -i  \frac{\a'}{2}\left[ {\rm tr} [R
\,\bg^{-1}\d \bg] - {\rm tr} [F \,\bh^{-1}\d \bh]\right]\Big)=0
~.} In the next section, we will write down explicit deformations
that satisfy the above equations for the $T^2$ bundle over $K3$
flux background.

\newsec{$T^2$ bundle over $K3$ solution}

The metric of the $T^2$ bundle over $K3$ solution \refs{\FY,
\BBFTY} has the form \eqn\metric{\eqalign{ds^2 &= e^{2 \phi}
ds_{K3}^2 + \left(dx+ \a_1\right)^2 + \left( dy + \a_2\right)^2\cr
&=e^{2 \phi} ds_{K3}^2 + |dz^3 + \a|^2 ~,}} where $\th=dz^3 + \a$
is a $(1,0)$-form and $\a = \a_1 + i \a_2$. The twisting of the
$T^2$ is encoded in the two-form defined on the base $K3$
\eqn\twoform{\om=\om_1+\,i\,\om_2=d\th=\om_S^{(2,0)}+\om_A^{(1,1)}~,}
which is required to be primitive \eqn\omprim{\om\w J_{K3} = 0~,}
and obeys the quantization condition \eqn\bv{\tom_i = {\om_i\over
2\pi\sqrt{\a'}} \in H^2(K3,\IZ)~.}  With this metric ansatz, the
anomaly cancellation equation reduces to a highly non-linear
second-order differential equation for the dilaton $\phi$.
Importantly, a necessary condition for the existence of a solution
for $\phi$ is that the background satisfies the topological
condition \eqn\kfcon{ \int_{K3} \left(\|\tom_S\|^2 +
\|\tom_A\|^2\right) {J_{K3}\w J_{K3} \o 2!} +
\frac{1}{16\pi^2}\int_{K3}{\rm tr}\, F\w F = 24 ~.} If this
condition is satisfied, then the analysis of Fu and Yau \FY\
guarantees the existence of a smooth solution for $\phi$ that
solves the differential equation of anomaly cancellation.

\subsec{Equations for the moduli}

For expressing the constraint equations of the allowed
deformations, we first write down more explicitly the hermitian
metric.  Note that the conventions we follow here are that
$J_{a\bb}=ig_{a\bb}$ and $ds^2 = 2 g_{a\bb}dz^a d\bz^\bb$.  The
hermitian two-form can be expressed simply as \eqn\Jform{J =
e^{2\phi}J_{K3} + \frac{i}{2}\, \th\w {\bar \th}~,} and we write
the corresponding metric as \eqn\gab{g_{a\bb}=
\frac{1}{2}\left(\matrix{2\gs+ \B\Bs & \B\cr \Bs & 1}\right)~,}
where $\gs_{i\bj}=e^{2\p}g_{K3}$ is the base $K3$ metric with the
$e^{2\p}$ warp factor included, $\B=(\B_1,\B_2)$ is a column
vector with entries locally given by $\a=\B_1 dz^1 + \B_2 dz^2\,$,
and $\Bs=\B^\dg$.

An allowed deformation of the conformally balanced condition must
satisfy the requirement that the four-form \confbalvv\
\eqn\ttcb{\eqalign{\d \rho &= \|\O\|_J\left[J\w \d J - \frac{1}{8}
(J \w J) J^{mn}\d J_{mn}\right] \cr & =  J_{K3}\w \d J +
\frac{i}{2} e^{-2\p} \th\w \bth\w \d J
-\,\frac{1}{8}(e^{2\phi}J_{K3}\w J_{K3} + i J_{K3}\w\th\w\bth)
J^{mn}\d J_{mn}~,}} is $d$-closed.

As for the anomaly condition, we shall work with the constraint given in the form of \lcondc\
(with trivial gauge bundle) \eqn\anomtt{\pa\bpa\big(\d J -
i\frac{\ap}{2} {\rm tr}  [R \,\bg^{-1}\d \bg]\big)= 0~.} The
curvature term can be written out explicitly as \eqn\trrexp{{\rm
tr}[R\,\bg^{-1}\d \bg]=  i \left(\tR^{\bj i} \d J_{i\bj} +
\tR^{\bj 3} \d J_{3\bj} + \tR^{{\bar 3} i}\d J_{i {\bar 3}} +
\tR^{{\bar 3} 3}\d J_{3{\bar 3}}\right)~,} where
\eqn\traa{\tR^{\bj i} = - \gs^{-1} R' - \frac{1}{2} (g'^{-1}
\bpa\B) (\pa\Bs g'^{-1})~,} \eqn\trac{\tR^{\bj 3} =  \gs^{-1} R'
\B + \pa(\gs^{-1}\bpa\B) - \frac{1}{2}(\gsi\bpa\B)(\pa\Bs\,
\gsi)\B~,} \eqn\trca{\tR^{{\bar 3}i}=  \Bs \gsi R' - \bpa(\pa\Bs
\gsi) + \frac{1}{2} \Bs (\gsi \bpa\B)(\pa\Bs\, \gsi)~,}
\eqn\trcc{\tR^{{\bar 3}3}= - \Bs \gsi R' \B + \frac{1}{2} \Bs \gsi
\bpa\B(\pa\Bs \gsi) \B + \bpa (\pa\Bs\gsi)\B -
\Bs\pa(\gsi\,\bpa\B) - \pa\Bs(\gsi\bpa\B)~,} and $R'=
\bpa({\bar{\gs}}^{-1}\pa{\bar {\gs}})$ is the curvature tensor of
$K3$ with respect to the $g'$ metric. Note that the $\tR^{\bb a}$
are two-forms with components only on the coordinates of $K3$.

Below, we shall analyze the infinitesimal deformations of the
$T^2$ bundle over $K3$ model with trivial gauge bundle.  For this
type of model, the topological constraint \kfcon\ is satisfied
purely by the curvature of the $T^2$ twist.  (See section~5.2~in
\BBFTY\ for explicit examples.)  We shall discuss the variation of
the three components of the metric - the dilaton conformal factor,
the $K3$ base, and the $T^2$ bundle - separately below.  We will
show that the moduli given below satisfy both the conformally
balanced and anomaly cancellation condition.  For the trivial
bundle case, the hermitian Yang-Mills condition does not place any
constraint on the deformations.
Finally, we will also discuss the variation of the complex structure in this model.

\subsec{Deformation of the dilaton}

The dilaton is associated to the warp factor of the $K3$ base.
Thus, varying the dilaton corresponds to varying the local scale
of the $K3$. The deformation of the hermitian form due to the
variation of the dilaton is \eqn\dilv{\d J = 2\,\d \p\, e^{2\p}
J_{K3}~,} where $\d \p$ depends only on the $K3$ coordinates. This
is consistent with the dilaton variation condition $\d \phi =
(1/8) J^{mn} \d J_{mn}$ of equation \dilvar.  As for the
conformally balanced condition, it in fact does not place any
constraint on the dilaton. The metric variation \dilv\ when
inserted into \ttcb\ gives the four-form \eqn\dilrho{\d \rho =
e^{2\phi} \JK \w \JK\, \d \phi~,} which is indeed $d$-closed for
any real function $\d \phi$ on the base $K3$. Since the space of
real function is infinite-dimensional, the dimensionality of the
deformation space is also infinite if only the conformally
balanced condition is considered.

Imposing anomaly cancellation condition will however make the
deformation space finite.  Anomaly cancellation \anomtt\ imposes
the condition \eqn\dilanomv{\pa\bpa \left( \Big[2 e^{2\p} J_{K3}
-i \frac{\a'}{2} e^{-2\p} {\rm tr}[\bpa \B \w \pa \Bs
g^{-1}_{K3}] + 4 \,\bpa \pa \p \Big]\d \p\right) = 0 ~,} where we
have used \traa. The analysis of Fu and Yau \FY\ guarantees only a
one-paramater family of solutions parametrized by the
normalization \eqn\psinorm{A=\left(\int_{K3}\, e^{-8\p} {J_{K3}\w
J_{K3} \o 2!}\right)^{1/4}~,} as long as the topological condition
\kfcon\ is satisfied and also $A\ll 1$.  (See \BBFTY\ for a discussion of the physical implications of the $A \ll 1$ bound.)
The variation of the
dilaton can thus be parametrized by the value of
$A$.\foot{Rigorously, one should be able to show that there does
not exist a dilaton variation that satisfies \dilanomv\ and leaves
the normalization $A$ unchanged. Regardless, the
finite-dimensionality of the deformation space is ensured if one
assumes the elliptic condition required by Fu and Yau \FY\ to
solve the anomaly cancellation equation for $\phi$.}

\subsec{Deformations of the K3 metric}

The metric moduli of the $K3$ are associated with deformations of
the hermitian form $J_{K3}$ such that the curvatures of the $T^2$
bundle, $\om_i$ for $i=1,2$, remain primitive \omprim. This
implies that the allowed variation of $\d J_{K3}$ satisfies
\eqn\curvar{\om_i \w \d J_{K3} + \d \om_i \w
J_{K3}=0~,~~~~i=1,2~.} Hence, of the $20$ possible $h^{1,1}$
K\"ahler deformations of $K3$, only the subset that satisfies
\curvar\ is allowed.

First, consider the case where $\d \om_i =0$.  We then have the
condition \eqn\curvsp{\om_i \w \d J_{K3} =0~,}
 which must be satisfied locally at
every point on $K3$.  With the curvature form $\om$ containing a
$(1,1)$ part, \curvsp\ is a very strong condition that in general
can only be satisfied by a variation proportional to the hermitian
form, $\d \JK \sim \JK$.  But this would then be the modulus
identified above as associated with the dilaton \dilv.

More generally, we can have $\d \om_i  = i \pa \bpa f_i$, where
$f_i$ for $i=1,2$ are functions on the base $K3$. This form of $\d
\om_i $ is required so that the variation does not change the
$H^2(K3)$ integral class of $\om_i$ as required by the
quantization of \bv. Let $\d J_{K3}= \eta \in H^{1,1}(K3)$ and not proportional to $J_{K3}$, then
the variation \curvar\ corresponds to
\eqn\curvart{\eqalign{0&=\om_i \w  \eta + i \pa \bpa f_i \w J_{K3}
\cr & = (f'_i - \Delta f_i ) \frac{\JK \w \JK}{2}~.}} Here, we
have replaced $\om_i \w \eta = f'_i \frac{\JK \w \JK}{2}$ noting
that the exterior product of two $(1,1)$-forms on the base must be
a function times the volume form of the $K3$.  Now, the sufficient
condition that a solution for $f_i$ exists is that
\eqn\curvari{\int_{K3} f'_i \frac{\JK \w \JK}{2} = \int_{K3} \om_i
\w \eta  = 0~.}   But this is related to the requirement that the
intersection numbers are zero. The intersection numbers of $K3$
are defined to be \eqn\interK{d_{IJ}= \int_{K3} \tom_I \w
\tom_J~,} where $\tom_I$, $I=1,\ldots,22\,$, denotes a basis of
$H^2(K3,\ZZ)$. The matrix $d_{IJ}$ is the metric of the even
self-dual lattice with Lorentzian signature $(3,19)$ given by
\eqn\selfdual{ (-E_8) \oplus (-E_8) \oplus \left(\matrix{0&1\cr
1&0}\right)\oplus \left(\matrix{0&1\cr
1&0}\right)\oplus\left(\matrix{0&1\cr 1&0}\right)~,} where
\eqn\Eeight{ E_8=\left( \matrix{ \hfill 2&\hfill 0&\hfill
-1&\hfill 0&\hfill 0&\hfill 0&\hfill 0&\hfill0\cr
\hfill0&\hfill2&\hfill0&\hfill-1 &\hfill 0&\hfill 0&\hfill
0&\hfill 0\cr \hfill -1&\hfill 0&\hfill 2&\hfill -1&\hfill
0&\hfill 0&\hfill 0&\hfill 0\cr \hfill 0&\hfill -1&\hfill
-1&\hfill 2&\hfill -1&\hfill 0&\hfill  0&\hfill 0\cr \hfill
0&\hfill  0&\hfill 0&\hfill -1 &\hfill 2&\hfill -1& \hfill 0&
\hfill 0\cr
 \hfill 0& \hfill 0& \hfill 0& \hfill 0&\hfill -1&\hfill 2&\hfill
 -1&\hfill  0\cr
 \hfill 0& \hfill 0&\hfill  0&\hfill  0&\hfill 0&\hfill -1&\hfill
 2&\hfill -1 \cr
\hfill  0&\hfill  0& \hfill 0& \hfill 0& \hfill 0& \hfill 0&\hfill
-1&\hfill 2 }\right)~,} is the Cartan matrix of $E_8$ Lie algebra.
Thus we see that a variation of $\d \JK = \eta$ is allowed as long
as the intersection numbers of $\eta$ with $\om_i$ are zero. This
implies at least that $\eta \neq \om_1\, ,\, \om_2$.

The above variations of the K\"ahler form on the $K3$ require the
metric variations \eqn\Kmetv{\eqalign{\d J & = e^{2\p} \eta +
\frac{i}{2}(\d\th \w \bth + \th \w \d\bth) + 2\,\d \p\, e^{2\p}
J_{K3}~,\cr \d \rho & = \frac{i}{2} \left( \th\w \bth \w \eta +
\JK\w (\d\th \w \bth + \th \w \d\bth)\right)+e^{2\phi} \JK \w \JK
\,\d \phi ~,}} where $\d\th = -i\,\pa(f_1+ i f_2)$.  One can check
that the above $\d \rho$ is closed when \curvart\ is satisfied.
We note that the additional variation of the dilaton in \Kmetv\ is
needed in order to satisfy the anomaly condition.  With it, the
analysis of Fu and Yau \FY\ then guarantees the existence of a
solution for $\d \p$ for each consistent pair $(\eta, \d \th)$.
Therefore, $\d J$ variations in \Kmetv\ satisfying \curvari\  are
indeed moduli.

\subsec{Deformation of the $T^2$ bundle}

We now consider the variation of the size of the $T^2$ bundle.
This is an allowed variation of the conformally balanced condition
since the metric variation \eqn\torv{\d J = \frac{i}{2}\,\e\,
\th\w\bth~,} results in the closed four-form \eqn\torrho{\d \rho =
\e\,(-\frac{1}{4} e^{2\p} \JK\w\JK + \frac{i}{4} \th\w\bth\w
\JK)~,} where $\e$ is a constant infinitesimal parameter.  But we
must also check the anomaly condition. The variation of the
curvature term can be calculated using  \traa-\trcc\ and we obtain
\eqn\ttworr{{\rm tr}  [R \,\bg^{-1}\d \bg] = \frac{1}{2}\,\e\,{\rm tr}
[\bpa \B\w \pa \B^* g'^{-1}]~.}
The anomaly condition \anomtt\ therefore becomes
\eqn\ttanom{\eqalign{0&=i\,\pa\bpa(\d J -
i\frac{\ap}{2}\, {\rm tr} [R \,\bg^{-1}\d \bg]) \cr & =
-\frac{1}{2}\,\e\,\pa\bpa (\th\w\bth - \frac{\ap}{2}\,{\rm tr}
[\bpa \B\w \pa \B^* g'^{-1}]) \cr &= \frac{1}{2}\,\e\,(\|\om\|^2
\frac{\JK^2}{2} + \frac{\ap}{2}\,\pa\bpa\,{\rm tr} [\bpa \B\w \pa
\B^* g'^{-1}])~,}} but this can not hold true. To see this, we can
integrate the last line over the base $K3$.  The first term gives
a positive contribution while the second term integrates to zero.
Here, we have used the fact that the two-form ${\rm tr} [\bpa \B\w
\pa \B^* g'^{-1}]$ in the second term is well-defined and has
dependence only on the base $K3$ as was shown in \FY\ (see Lemma
10 on page 11). Thus, the size of the torus can not be
continuously varied as it is fixed by the anomaly condition.

With the size of the torus fixed, it is evident that there can not
be any overall radial moduli $\d J = \e\,J$ for this model, as has
also been noted previously in \refs{\BBDP, \locu, \BT}.  Actually,
it is true in general that the anomaly cancellation forbids an
overall constant radial modulus for any heterotic compactification
with non-zero $H$-flux. The reason is simply that ${\rm tr} [R\w
R]$ is invariant under constant scaling of the metric since the
Riemann tensor, $R_{mn}{}^p{}_q\,$, is scale invariant. However
$dH= 2i\,\pa\bpa J$ depends on $J$ and can not be scale invariant.
Hence, the overall scale is not a modulus.

To summarize, the $T^2$ bundle over $K3$ model has a dilaton
modulus and also moduli associated with the K\"ahler moduli of the
base $K3$.  The number of moduli in particular depends on the
curvature of the $T^2$ twist, $\om$.  The size of the $T^2$ is
however fixed and hence there is no overall radial modulus in the
model.

\subsec{Fixing the complex structure}

We have mostly taken the complex structure to be fixed in
analyzing the moduli.  But for the $T^2$ bundle over $K3$
solution, the complex structures are rather transparent and we can
describe how they can be fixed.  To begin, the complex structures
are simply those on the $K3$ plus that on the $T^2$.  For the
$T^2$, its complex structure determines the integral first Chern
class quantization condition \bv\ for $\om_1$ and $\om_2$.  For an
arbitrary torus complex structure $\tau=\t_1 + i\,\t_2\,$, the
quantization conditions depend on $\tau$ and takes the form
\eqn\complexfixed{\frac{1}{2\pi\sqrt{\a'}}\int_\G (\om_1 - \frac{\t_1}{\t_2} \om_2) \in
\IZ~,\qquad\qquad \frac{1}{2\pi\sqrt{\a'}\,\t_2}\int_\G \om_2 \in
\IZ~,}
where $\G\in H_2(K3,\ZZ)$ is any two-cycle on $K3$.  Therefore, fixing $\om = \om_1 + i \om_2$ effectively fixes $\t$.  And even if we were to allow $\om$ to vary infinitesimally, the complex structure integrability condition $\om=\om_1 + i \,\om_2 \in \Lambda^{(2,0)}(K3) \oplus \Lambda^{(1,1)}(K3)$ and the topological condition \kfcon\ must be imposed.  All together, these strong conditions generically fix the $T^2$ complex structure moduli.\foot{That the $T^2$ complex structures are fixed has also be noted from the gauged linear sigma model point of view in \adel.}  Note also that the condition $\om\in H^{(1,1)}(K3,\ZZ)=H^{(1,1)}(K3)\cap H^2(K3,\ZZ)$ also strongly constrains the complex structure of the $K3$ since the dimension of $H^{(1,1)}(K3,\ZZ)$ do vary with the complex structure of $K3$.

The complex structures of $K3$ can also be fixed if the $T^2$
twist $\om$ contains a $(2,0)$ self-dual part, $\om^{(2,0)}=k\,
\O_{K3}$, which up to a constant $k$ must be proportional to the
holomorphic $(2,0)$-form of $K3$. The above mentioned quantization
condition for the $(2,0)$ part then takes the form (for $\tau=i$)
\eqn\period{\frac{k}{2\pi\sqrt{\alpha'}}\int_\G \O_{K3} \in \IZ~,}
which defines the periods of the holomorphic $(2,0)$-form on the
$K3$.  These periods specify the complex structures chosen on
$K3$, and the quantization condition thus fixes the complex
structures on $K3$.

\newsec{Conclusions and Open Questions}

In this paper, we have derived the defining equations for the
local moduli of supersymmetric heterotic flux compactifications.
The defining equations were derived by performing a linear
variation of the supersymmetry constraints obeyed by such
compactifications. We further analyzed the corresponding geometric
moduli spaces and discussed the particular example of a $T^2$
bundle over $K3$ in detail. This $T^2$ bundle over $K3$ solution
is special in that in it is dual to M- or F-theory on $K3\times
K3$.  Notice that under infinitesimal deformations, the manifold
$K3\times K3$ remains $K3\times K3$.  Thus, the corresponding
heterotic $T^2$ bundle over $K3$ dual must also be locally unique;
that is, it remains a $T^2$ bundle over $K3$ under infinitesimal
variation.

In much of our analysis, we have set the gauge bundle to be
trivial.  For the $T^2$ bundle over the $K3$ case, the
non-trivial, non-$U(1)$ bundle are simply the stable bundles on
$K3$ lifted to the six-dimensional space.  The moduli space then
corresponds to the space of $K3$ stable bundle.  The dimension of
this moduli space $M$ is given by the Mukai formula \Mukai
\eqn\mukai{{\rm dim}\, M = 2r\,c_2(E) -(r-1) c_1^2(E) -2 r^2 +
2~,} where $r$ is the rank of the bundle (i.e.~the dimension of
the fiber), and $(c_1(E), c_2(E))$ are the first and second Chern
number of the gauge bundle $E$.  It would be interesting to
understand the moduli space of stable gauge bundle in general.

There are a number of interesting open questions. First, in our
analysis we have kept for simplicity the complex structure fixed.
It is well known that for Calabi--Yau compactifications the moduli
space is a direct product of complex structure and K\"ahler
structure deformations.  For non-K\"ahler manifolds with torsion,
this likely is not the case and it would be interesting to allow
for a simultaneous variation of the complex structure and the
hermitian form.

It would be interesting to analyze the geometry of the moduli space
and to determine if powerful tools such as the well known ``special
geometry'' of Calabi--Yau compactifications \refs{\Strominger}\ can
be derived in this case.

Furthermore, counting techniques for moduli fields need to be
developed and we expect that the number of moduli can be
characterized in terms of an index or some topological invariants of the manifold.

Finally, it would be interesting to analyze the moduli space from
the world-sheet approach using the recently constructed gauged
linear sigma model \refs{\adel}. Moduli fields will correspond to
the marginal deformations of the IR conformal field theory.

\bigskip\bigskip\bigskip

\centerline{\bf Acknowledgements}
\medskip
 We would like to thank A.~Adams, K.~Becker, S.~Giddings, J.~Lapan,
 J.~Sparks,
 E.~Sharpe, A.~Subotic, V.~Tosatti, D.~Waldram, M.-T.~Wang, P.~Yi, and especially J.-X.~Fu for helpful discussions.
 We thank the 2006 Simons Workshop at YITP Stony Brook for hospitality where part of this work
 was done.  M.~Becker would like to thank members of the
 Harvard Physics Department for their warm hospitality during the final
 stages of this work.  The work of M.~Becker is supported by NSF grants
 PHY-0505757, PHY-0555575 and the University of Texas A\&M. The work
 of L.-S.~Tseng is supported in part by NSF grant DMS-0306600 and Harvard University.  The work
 of S.-T.~Yau is supported in part by NSF grants DMS-0306600, DMS-0354737, and DMS-0628341.

\bigskip\bigskip\bigskip

\centerline{\bf Appendix}

\medskip

We summarize our notation and conventions.

\item{$\bullet$} Our index conventions are as follows:
$m,n,p,q,\ldots$ denotes real six-dimensional coordinates, $a, b,
c, \ldots$ and $\ba, \bb, \bc, \ldots$ denote six-dimensional
complex coordinates, and $i,j,k,\ldots$ and ${\bar i}, {\bar j},
{\bar k}, \ldots$ denote four-dimensional complex coordinates on
the base $K3$.

\item{$\bullet$} The gauge field $A_m$ and field strength $F_{mn}$
take values in the $SO(32)$ or $E_8\times E_8$ Lie-algebra with
the generators being anti-hermitian.

\item{$\bullet$} The Riemann tensor is defined as follows
$$
R_{mn}{}^p{}_q=\pa_m \G_n{}^p{}_q-\pa_n \G_m{}^p{}_q +
\G_m{}^p{}_r\, \G_n{}^r{}_q-\G_n{}^p{}_r\,\G_m{}^r{}_q~.$$ With a
hermitian metric $g$ with components $g_{a\bb}$, we write the
hermitian curvature two-form as $R=\bpa[\bg^{-1}\pa \bg]=\bpa[(\pa
g) g^{-1}]$ where $\bg$ is the transposed of $g$ with components
$g_{\bb a}$.  Explicitly, in components, we write
$$R_{\ba b}{}^c{}_d = \bpa_\ba[g^{c\bd}\pa g_{\bd d}]=\bpa_\ba[(\pa_b g_{d\bd})g^{\bd c}]~.$$

\item{$\bullet$} We follow the convention standard in the
mathematics literature for the Hodge star operator.  For example,
$(\star H)_{mnp}= {1\o 3!}\,H_{rst}\,\epsilon^{rst}{}_{mnp}$ with
$\epsilon_{mnprst}$ being the Levi-Civita tensor.

\item{$\bullet$} We use the definition for $\|\O \|_J^2$:
$$
\O\,\w\, \star\, {\bar \O} = \|\O\|_J^2 {J^3 \o 3!}~ .
$$

\item{$\bullet$} For a vector field, $v=v^m \pa_m$, the interior
product acting on a $p$-form with components $\a_{m_1m_2\ldots
m_p}$ is just
$$(i_v \a)_{m_2 m_3 \ldots m_p} = v^{m_1}\a_{m_1m_2\ldots m_p}~.$$

\item{$\bullet$} Given a hermitian form $J$, the adjoint of the
Lefschetz operator $\La$ acting on a $p$-form with components
$\a_{m_1m_2\ldots m_p}$ is
$$(\La \,\a)_{m_3 m_4 \ldots m_p} = \frac{1}{2!} J^{m_1m_2} \a_{m_1 m_2 m_3 m_4\ldots m_p} ~.$$

\listrefs

\end